\documentclass[english,aps,prd,nofootinbib,showkeys,preprint,floatfix]{revtex4}
\pdfoutput=1
\usepackage[utf8]{inputenc}
\usepackage[english]{babel}
\usepackage{amsmath}
\usepackage{amssymb}
\usepackage{graphicx}
\usepackage{multirow} % To use multirow command in tables
\usepackage[colorlinks=True]{hyperref}
\usepackage{lmodern}

\newcommand{\AddrUsM}{
{\it Universidad T\'ecnica Federico Santa Mar\'ia and Centro Cient\'ifico Tecnol\'ogico de Valparaiso,\\
    Casilla 110-V, Valparaiso, Chile.}}

\newcommand{\AddrUdeA}{%
 Instituto de F\'\i sica, Universidad de Antioquia,\\
 Calle 70 No. 52-21, Medell\'in, Colombia. }

\begin{document}

%\preprint{XXXX}  

\title{Fermion dark matter from SO(10)}

\author{Carolina Arbel\'aez}\email{carolina.arbelaez@.usm.cl}
\affiliation{\AddrUsM} 
\author{Robinson Longas}\email{robinson.longas@udea.edu.co}
\affiliation{\AddrUdeA} 
\author{Diego Restrepo} \email{restrepo@udea.edu.co}
\affiliation{\AddrUdeA}
\author{Oscar Zapata} \email{oalberto.zapata@udea.edu.co}
\affiliation{\AddrUdeA}


\begin{abstract}
We construct and analyze nonsupersymmetric $\text{SO}(10)$ standard model
extensions which explain dark matter (DM) through the fermionic Higgs
portal.
In these $\text{SO}(10)$-based models the DM particle is naturally stable since a $Z_2$ discrete symmetry, the matter parity, is left at the end of the symmetry breaking chain to the standard model. 
Potentially realistic models contain the $\bf{10}$ and $\bf{45}$
fermionic representations from which a neutralino-like mass matrix with arbitrary mixings can be obtained.
Two different $\text{SO}(10)$ breaking chains will be analyzed in light of
gauge coupling unification: the standard path $\text{SU}(5)\times
U(1)_{X}$ and the left-right symmetry intermediate chain. 
The former opens the possibility of a split supersymmetric-like
spectrum with an additional (inert) scalar doublet, while the later requires
additional exotic  scalar representations associated to the breaking
of the left-right symmetry. 
\end{abstract}

\maketitle

\section{Introduction}
In view of the lack of signals of new physics in strong production at
the LHC, the naturalness criterion as a guide to build extensions of
the standard model has been losing priority in favor of other
theoretical and phenomenological motivations.  

Split supersymmetry (split-SUSY)~\cite{ArkaniHamed:2004fb,Giudice:2004tc,ArkaniHamed:2004yi} for
example, gives up the explanation of the hierarchy problem while
keeping the other main virtue of the minimal supersymmetric standard
model: the connection between gauge coupling unification (GCU) and
viable dark matter (DM) candidates without the imposition of any \emph{ad hoc} discrete
symmetry. 
In fact, the discrete symmetry required to avoid fast proton decay in
supersymmetry can be embedded in an anomaly-free gauge symmetry (see
for instance
\cite{Ibanez:1991hv,Ibanez:1991pr,Babu:2003qh,Dreiner:2005rd,Paraskevas:2012kn})
in order to avoid quantum gravitational effects which would violate
it~\cite{Krauss:1988zc,Banks:1991xj,Mambrini:2015sia}.  
If in addition, the emerging discrete symmetry also forbids lepton ($L$)
and baryon  number ($B$) violation in the superpotential, the lightest
supersymmetric particle is rendered stable with the potential to be a
good dark matter candidate~\cite{Sierra:2009zq,Dreiner:2012ae,Florez:2013mxa}.

One straightforward possibility arises if split-SUSY is built
in the framework of $\text{SO}(10)$-GUT~\cite{Wang:2015mea}.  
If we break the $\text{U}(1)_{B-L}$ subgroup of $\text{SO}(10)$ by the vacuum expectation
values (VEVs) of fields with even $B-L$, then the discrete
symmetry $P_{M}=(-1)^{3(B-L)}$, known as matter
parity~\cite{Dimopoulos:1981zb,Dimopoulos:1981dw}, is preserved. In such a case,
both the proton and dark matter stability are guaranteed at the
renormalizable level. 

It is interesting to stress that this possible DM stability
explanation is independent of supersymmetry and can also happen when
the Standard Model (SM) is embedded in
$\text{SO}(10)$.~\cite{Fritzsch:1974nn}\footnote{In the minimal dark matter
  scenario \cite{Cirelli:2005uq}, the DM candidate is either a scalar septuplet or fermion
  quintuplet of $\text{SU}(2)_L$ and its stability is guaranteed by the SM gauge
  symmetry.}
Being a rank 5 group it contains an additional $U(1)_X$ subgroup,
apart from the SM group, and if it is spontaneously broken by a scalar
field $S$ having a nonzero $U(1)_X$ charge $X_S$ with $X_S=0$ (mod
$N$) and $N\ge2,3,...,$ then a remnant $Z_N$ symmetry is expected to
be present even at low energies.  
This makes $\text{SO}(10)$ to be a promising group to explain the origin of the DM
stability \cite{Kadastik:2009dj}.
Hence, some simplified DM models have been analyzed in light of the
stability from $\text{SO}(10)$. 
In particular, the scalar doublet and singlet dark matter have been
studied in \cite{Kadastik:2009dj,Kadastik:2009cu}, the triplet
(singlet) fermion DM was considered in
Refs.~\cite{Frigerio:2009wf,Parida:2010jj} (\cite{Mambrini:2015vna}),
whereas the radiative seesaw model was analyzed in
Ref.~\cite{Parida:2011wh}.

In the last reference, it was also shown that a robust GCU can be
obtained in $\text{SO}(10)$
when the set of low energy fields emerging from the even $B-L$
fermionic representations $\mathbf{10}_F$ and $\mathbf{45}_F$ matches the
particle content of split-SUSY with one additional scalar doublet.
In this way, the spectrum matches exactly the low energy particle
content of partial split-SUSY (PSS)~\cite{Diaz:2006ee}. 
To our knowledge, this minimum set of fields was first proposed in Ref.~\cite{Ma:2005he}.

In this paper, we show that with the Yukawa couplings of the low energy
fields associated to the mixing of $\mathbf{10}_F$ and
$\mathbf{45}_F$, through the Higgs field in $\mathbf{10}_H$, we can
obtain a neutralino-like mass matrix but with different mixings
compared to the usual gauginos and Higgsinos. 
In this way, our framework automatically gives an explanation for the
origin of the \emph{ad hoc} discrete symmetry of simplified fermion dark
matter models connected to the Higgs portal~\cite{Patt:2006fw}. 
In  particular, we formulate $\text{SO}(10)$ realizations of the
singlet-doublet fermion dark matter (SDFDM) model~\cite{ArkaniHamed:2005yv,Mahbubani:2005pt,D'Eramo:2007ga,Enberg:2007rp} and the doublet-triplet
fermion dark matter (DTFDM) model~\cite{Dedes:2014hga}.

Scalar and vector DM naturally make use of the Higgs portal through the invariant Higgs mass factor $H^\dagger H$  
whereas fermion DM requires an ultraviolet (UV)
realization, via a scalar or a vector mediator, of the dimension-5 terms
$\bar{F}FH^\dagger H$ and $\bar{F}\gamma_5FH^\dagger H$ \cite{Djouadi:2011aa}.
The singlet fermion DM model \cite{Kim:2006af,Kim:2008pp,Baek:2011aa} is a UV realization of the fermionic Higgs portal \cite{Patt:2006fw,LopezHonorez:2012kv,Fedderke:2014wda,Freitas:2015hsa} with an additional 
singlet scalar (to be mixed with the Higgs) as the mediator. 
On the other hand, for those simplified models where the DM particle 
is a mixture of either singlet and doublet fermions or doublet and 
triplet fermions, the mediator particle is the Higgs itself. 
Thus the SDFDM   
and DTFDM models are 
two of the simplest fermionic DM models where the Higgs portal 
is open without additional scalar degrees of freedom.  
This, along with the Higgs boson discovery and the lack of signals of new physics 
in strong production at the LHC, make of these simplified fermion DM models 
(where the production of new particles is only through electroweak processes) 
a realistic and promising solution to the DM puzzle.

With the PSS-like spectrum as in \cite{Parida:2011wh}, we
revisit the GCU with emphasis in a scenario where $\text{SO}(10)$ breaks to
the SM through the $\text{SU}(5)\times \text{U}(1)_{X}$ chain.  

Finally, we explore the possibility to have a correct non-SUSY $\text{SO}(10)$ GCU with 
another  kind of spectrum  in the ballpark of the
$O(1)$ TeV. 
Since the triplet fermion DM model only requires one additional colored octet at
some high scale to achieve GCU~\cite{Frigerio:2009wf,Parida:2010jj},
we will focus in the case when only the singlet-doublet fermions
contains the DM candidate.
There exists much literature which already discusses GCU for the breaking
chain of $\text{SO}(10)$ containing the left-right
(LR) symmetric gauge group with remnant gauge $U(1)_{B-L}$ symmetry
\cite{Siringo:2012bc,Mahbubani:2005pt,Lindner:1996tf,Dev:2009aw,Bertolini:2009es,DeRomeri:2011ie,Arbelaez:2013hr}. 
We will check specifically if SDFDM is compatible with a low LR intermediate symmetry breaking, adding at this level a few extra particle content
imposed to pass some specific phenomenological constraints.

In the next section,
we present the minimal $\text{SO}(10)$ setup to realize the fermionic DM Higgs portal.
In the Sec. \ref{sec:GCU}, we analyze the GCU for some models which 
successfully constitute a fermionic DM Higgs portal realization, and interesting 
configurations will be explored. Finally, in Sec. \ref{sec:conclusions} we 
present our conclusions.

\section{Fermion DM from $\text{SO}(10)$}
\label{sec:simplifiedDM}
The split-SUSY scenario demands that the supersymmetric partners of the leptons and quarks along the second Higgs doublet stay at an intermediate scale $M_S\gg 1$ TeV, whereas the first Higgs doublet, Higgsinos and gauginos remain at low energies \cite{ArkaniHamed:2004fb,Giudice:2004tc,ArkaniHamed:2004yi}. Therefore, any nonsupersymmetric version of such a scenario involves the following particle content: a hyperchargeless singlet Weyl fermion $N$, two $\text{SU}(2)_L$-doublets Weyl fermions $\chi$, $\chi^c$ with opposite hypercharge $Y=\pm 1/2$, a hyperchargeless $\text{SU}(2)$-triplet Weyl fermion $\Sigma$ and a color octec Weyl fermions  $\Lambda$ with $Y=0$. To generate this particle spectrum from $\text{SO}(10)$, we choose its $P_M$-even vector ${\bf 10}_F$  and adjoint ${\bf 45}_F$ fermion representations~\cite{Parida:2011wh}. Concretely, $N,\Sigma$ and $\Lambda$ belong to the adjoint representation, and the Weyl doublets are in the vectorial one. 
As usual the SM fermions are in the $P_M$-odd spinorial ${\bf 16}_a$ representation ($a=1,2,3$ is the family index), while the Higgs field is assigned to the fundamental representation ${\bf 10}_H$. In this way, the matter parity guarantees the stability of the dark matter particle, which is a mixture of all the $P_M$-even neutral colorless fermions in the spectrum.

The most general $\text{SO}(10)$ invariant Lagrangian contains the following Yukawa terms
\begin{align}
\label{eq:lagrangian}
-\mathcal{L} & \supset  Y{\bf 10}_F {\bf 45}_F {\bf 10}_H + M_{\mathbf{45}_F}{\bf 45}_F{\bf 45}_F + M_{\mathbf{10}_F}{\bf 10}_F{\bf 10}_F. 
\end{align}
To break the mass degeneracy within ${\bf 10}_F$ and  ${\bf 45}_F$ multiplets and at the same time generate low scale masses for the nonstandard fermions it is enough to consider the additional scalar representations $ {\bf 45}_H,\, {\bf 54}_H$ and ${\bf 210}_H$ \cite{Parida:2011wh,Parida:2010jj}. Concretely, the Lagrangian involving the $\mathbf{10}_F$ and $\mathbf{45}_F$ mass terms reads~\cite{Malinsky:2007qy,Parida:2011wh,Parida:2010jj}
\begin{align}
  \mathcal{L}_{\mathbf{10}_F+\mathbf{45}_F}^{\text{mass}}=&\,
{\bf 10}_F \left( M_{\mathbf{10}_F}+h_e' \left\langle \mathbf{54}_H \right\rangle\right) {\bf 10}_F+{\bf 45}_F \left( M_{\mathbf{45}_F}+h_e \left\langle \mathbf{54}_H \right\rangle+h_p \left\langle \mathbf{210}_H \right\rangle \right) {\bf 45}_F\,.
\end{align}
Since $\mathbf{210}_H$ have three singlets, while $\mathbf{54}_H$ has only one, the full set of masses  are
\begin{align}
  m(1,2,1/2)=&\, M_{\mathbf{10}_F}+\frac{3h_e'}{2}\left\langle \mathbf{54}_H \right\rangle, \nonumber \\
  m(3,1,-1/3)=&\, M_{\mathbf{10}_F}-h_e'\left\langle \mathbf{54}_H \right\rangle, \nonumber \\
  m(3,1,2/3)=&\, M_{\mathbf{45}_F}+{\sqrt 2}h_p{\left\langle \mathbf{210}_H \right\rangle_2 \over 3} -
2h_e{\left\langle \mathbf{54}_H \right\rangle\over {\sqrt {15}}}, \nonumber \\
m(3,2,1/6)=&\, M_{\mathbf{45}_F} + h_p{\left\langle \mathbf{210}_H \right\rangle_3\over 3} +h_e{\left\langle \mathbf{54}_H \right\rangle\over {2
  \sqrt{15}}},\nonumber \\
m(3,2,-5/6)=&\, M_{\mathbf{45}_F} - h_p{\left\langle \mathbf{210}_H \right\rangle_3\over 3} +h_e{\left\langle \mathbf{54}_H \right\rangle\over {2
  \sqrt{15}}},\nonumber \\  
m(1,1,0)=m(1,1,1)=&\, M_{\mathbf{45}_F} +\sqrt {2/3}h_p \left\langle \mathbf{210}_H \right\rangle_1 +
\sqrt {3/5}h_e\left\langle \mathbf{54}_H \right\rangle,\nonumber \\ 
m^{\prime}(1,1,0)=&\, M_{\mathbf{45}_F} +{2\sqrt {2}\over 3} h_p\left\langle \mathbf{210}_H \right\rangle_2 -
{2\over \sqrt{15}}h_e\left\langle \mathbf{54}_H \right\rangle,\nonumber \\ 
m(8,1,0)=&\, M_{\mathbf{45}_F} -{{\sqrt {2}}\over 3} h_p\left\langle \mathbf{210}_H \right\rangle_2 -
{2\over {\sqrt{15}}}h_e \left\langle \mathbf{54}_H \right\rangle,\nonumber \\ 
m(1,3,0)=&\, M_{\mathbf{45}_F} -\sqrt{2\over 3} h_p\left\langle \mathbf{210}_H \right\rangle_1 +\sqrt{3\over
  5}h_e\left\langle \mathbf{54}_H \right\rangle.\nonumber
%\label{masses}
\end{align}
Solving in terms of $M_{D}=m(1,2,1/2)$, $M_{\Lambda}=m(8,1,0)$, $M_{\Sigma}=m(1,3,0)$, and $M_N=m^{\prime}(1,1,0)$, we have that all the other masses are of order $M_{\mathbf{10}_F},\,M_{\mathbf{45}_F}\sim m_G$, except for
\begin{align}\label{eq:mTq}
  M_T= m(3,1,2/3)=\left( M_{\Lambda}+2 M_N \right)/3\,.
\end{align}
Therefore the fermion spectrum (FS) at low-intermediate energies can involve $N,\chi,\Sigma,\Lambda$ and/or $T$. Namely we have the following possibilities for the fields belonging to $\mathbf{45}_F$ having arbitrary masses, i.e., their masses are free parameters:
\begin{align}
\label{eq:LE}
\text{FS}_{\mathbf{45}_F}\, \text{I}:&\qquad\Sigma,\,\Lambda,\qquad\text{with}\qquad M_{N},\,M_{T}\sim M_{G},\\
\label{eq:NE}
\text{FS}_{\mathbf{45}_F}\, \text{II}:&\qquad N, \Sigma,\qquad\text{with}\qquad M_{\Lambda},\,M_{T}\sim M_{G},\\
%m_{\Lambda}\sim m_{G}\to& m_{T}\sim m_{G},& \qquad\text{free parameters: }& m_{N},m_{\Sigma}\\
\label{eq:LNE}
\text{FS}_{\mathbf{45}_F}\, \text{III}:&\qquad N,\Lambda,\, \Sigma,\, T.
%m_{T}=&\left( m_{\Lambda}+2 m_N \right)/3\,, \qquad\text{free parameters: }&m_{\Lambda}, m_{N},m_{\Sigma} \,.
\end{align}
It is worth mentioning that for the fermion spectrum~(\ref{eq:LNE}) the VL up-type quark $T$ is required due to  Eq.~(\ref{eq:mTq}). 
However, if a second innocuous $\mathbf{45}_F$ is introduced with the corresponding singlet $m'(1,1,0)$ having an arbitrary mass such a VL quark can be removed from the spectrum~(\ref{eq:LNE}) ($M_N,M_T\sim m_G$),  leading to another spectrum comprising the spectrum~(\ref{eq:LE}) plus a new singlet denoted again as $N$:
\begin{align}
\label{eq:LNE2}
\text{FS}_{\mathbf{45}_F}\, \text{IV}:&\qquad N,\Lambda,\, \Sigma.
%m_{T}=&\left( m_{\Lambda}+2 m_N \right)/3\,, \qquad\text{free parameters: }&m_{\Lambda}, m_{N},m_{\Sigma} \,.
\end{align}

%Let's note that there are other $SO(10)$ representations where the scalar and fermion fields can be accomodated. For instance, the ${\bf 45}_F$ representation can be exchanged with the ${\bf 54}_F$, the ${\bf 10}_F$ with the ${\bf 120}_F$, or the ${\bf 10}_H$ with the ${\bf 120}_H$.  
The $\text{SO}(10)$ breaking leads to the effective DM Yukawa Lagrangian for the fermion spectrum (\ref{eq:LNE}) with the pair $\chi,\,\chi^c,$\footnote{Here we use the additional scalar representations $ {\bf 120}_H$ and ${\bf 320}_H$, along the renormalization group equations, to generate a hierarchy between the four Yukawa couplings $y_i$ and $f_i$ at low energies. This implies that the Higgs doublet $H$ is a linear combination of the weak doublets present in $ {\bf 10}_H$, $ {\bf 120}_H$ and ${\bf 320}_H$. Another way to generate such a hierarchy is taking the ${\bf 10}_H$ as complex \cite{Bajc:2005zf}. In that case, to avoid the coupling of the SM fermions to ${\bf 10}_H^*$ an additional global $U(1)_{PQ}$ symmetry may be imposed leading to the axion as the DM candidate \cite{Bajc:2005zf, Babu:2015bna}. Because we are interested in WIMP fermion DM, we will not consider this case here.}
\begin{align}
\label{eq:lagrangianlow}
\mathcal{L}_{eff}  =&\,   M_{D}\chi^c\chi - \frac{1}{2}M_N NN - \frac{1}{2}M_\Sigma \Sigma\Sigma %- \frac{1}{2}M_\Lambda \Lambda\Lambda
\nonumber\\
&  - y_1H\chi^cN -y_2\tilde{H}\chi N + f_1H\epsilon \Sigma \chi^c  - f_2 \tilde{H}\epsilon \Sigma\chi +{\rm h.c.}
\end{align}		 
In this way, the opening of the Higgs portal through the $y_i$ and
$f_i$ terms allows the construction of the general scenario of
singlet-doublet-triplet fermion DM,  a neutralino-like scenario.
After the electroweak symmetry breaking the $y_i$ and $f_i$ terms induce a mixture between all the colorless $P_M$-even neutral fermions, and thus also a breaking of the mass degeneracy between
the neutral parts of the doublet fermions $\chi$ and $\chi^c$. 
It follows that the particle spectrum consists of four Majorana fermions and two Dirac charged fermions.
The neutral fermion mass matrix in the basis $\boldsymbol{\psi}^0=\left(N,\Sigma^0,\chi^{c\,0},\chi^{0}\right)^T$ reads
\begin{align}
\label{eq:Mchi}
  \mathcal{M}_{\boldsymbol{\psi}^0}=\begin{pmatrix}
 M_N         &   0       &-y\cos\beta v/\sqrt{2}& y\sin\beta v/\sqrt{2}\\
0 & M_\Sigma &  f\cos\beta'v/\sqrt{2} & -f\sin\beta'v/\sqrt{2}\\
-y\cos\beta v/\sqrt{2} &  f\cos\beta'v/\sqrt{2}  & 0            & -M_D\\
y\sin\beta v/\sqrt{2}& -f\sin\beta'v/\sqrt{2}&  -M_D                &  0  \\
\end{pmatrix},
\end{align}
while charged fermion mass matrix in the basis $\psi^+=(\Sigma^+,\,\chi^{+})^T$ and $\psi^-=(\Sigma^-,\,\chi^{c-})^T$ is given by
\begin{align}
\label{eq:MchiC}
  \mathcal{M}_{\boldsymbol{\psi}^\pm}=\begin{pmatrix}
 M_\Sigma         &   f\sin\beta'v   \\
f\cos\beta'v & M_D \\
\end{pmatrix}.
\end{align}
Here $y=\sqrt{y_1^2+y_2^2}$, $f=\sqrt{f_1^2+f_2^2}$, $\tan\beta=y_2/y_1$ and $\tan\beta'=f_2/f_1$. 
These mass matrices have the typical structure of the very well-known
neutralino and chargino mass matrices in the minimal supersymmetric
standard model (MSSM) \cite{Martin:1997ns}.  
Indeed, the supersymmetric case corresponds to the limit
$y=g'/\sqrt{2}$, $f=g/\sqrt{2}$ and $\tan\beta=\tan\beta'$.

It is worth mentioning the crucial role of the mixing terms in the neutral fermion sector. In the absence of them, the singlet DM would not couple to the SM particles thus leading a large relic abundance while the doublet DM would be excluded due to the coupling to the $Z$ gauge boson which gives rise to a spin-independent cross section orders of magnitude larger than present limits. The only limiting case that does not require the mixing terms is the triplet DM one. 

The present fermion particle spectrum was considered in Ref.~\cite{Carena:2004ha} with the aim of strengthen the first order electroweak phase transition in order to have a successful  electroweak baryogenesis. A neutralino-like mass matrix was also realized in fake split-SUSY~\cite{Benakli:2013msa} but with suppressed mixings  between fake gauginos and fake higgsinos. 

On the other hand, it is also possible to generate simpler DM
scenarios by assuming a mass hierarchy among the neutral fermions. For
$M_\Sigma\gg M_N,M_D$ the simplified model of SDFDM
~\cite{ArkaniHamed:2005yv,Mahbubani:2005pt,D'Eramo:2007ga,Enberg:2007rp}
is obtained\footnote{One alternative GUT scenario to have
  singlet-doublet fermion DM was presented in~\cite{Mahbubani:2005pt}.}, whereas the  DTFDM model \cite{Dedes:2014hga,Abe:2014gua,Freitas:2015hsa} emerges when $M_N\gg M_\Sigma,M_D$. Of course, the triplet fermion DM model \cite{Cirelli:2005uq} is also possible as long as $M_\Sigma\ll M_N,M_D$~\cite{Frigerio:2009wf,Parida:2010jj}.

The phenomenology of the  model in direct and indirect dark matter
detection experiments, and in colliders, is usually studied in the limits of
simplified fermion dark matter through the Higgs portal~\cite{Freitas:2015hsa,Abe:2014gua},  with emphasis in couplings which depart from SUSY
limit. In this way, the SDFDM has been  thoroughly studied 
in several works
\cite{ArkaniHamed:2005yv,Mahbubani:2005pt,D'Eramo:2007ga,Enberg:2007rp,Cohen:2011ec,Cheung:2013dua,Abe:2014gua,Restrepo:2015ura,Calibbi:2015nha,Freitas:2015hsa,Abdallah:2015ter}. 
The dark matter candidate is the lightest state coming from the mixing of the neutral component of the doublet and the neutral singlet. 
When the dark matter candidate is mainly singlet
(doublet) the relic density is in general rather large (small). 
In particular, a pure doublet has the proper relic density for
$M_{\chi}\sim1$ TeV~\cite{Mahbubani:2005pt,Cheung:2013dua,Chattopadhyay:2005mv}.  
The LHC phenomenology was analyzed in~\cite{Abe:2014gua}. Their conclusion, is that
the recast of the current LHC data is easier to evade, but the
long-rung prospects are promising, since the region $M_N,y_1v,y_2v\ll M_D$ could be 
probed up to $M_D\lesssim 600-700$~GeV for the 14-TeV run of the LHC with 
$3000\ \text{fb}^{-1}$. 

On the other hand, the phenomenology of the DTFDM has been
studied in~ \cite{Dedes:2014hga,Abe:2014gua,Freitas:2015hsa}.
The dark matter candidate is the lightest state coming from the mixing of the neutral components of the doublet and the triplet. In the low DM mass region, the relic density is properly satisfied in the range $0\leq (M_D,M_\Sigma)\leq400$ GeV and $0\leq (f,f')\leq1.5$. However this region is excluded due to the contribution of the new charged fermions to the Higgs diphoton decay \cite{Abe:2014gua}. For the high DM mass region, the expectations are analogous to the ones of the doublet or triplet fermion DM, where a large value for the DM mass is required.  
When the doublet is decoupled, the triplet fermion dark matter model is recovered with a mass of $\sim 2.7\ $GeV  to explain the correct
relic abundance~\cite{Cirelli:2005uq}. Therefore, its phenomenology at near-future colliders is quite limited~\cite{Abe:2014gua}.

\section{$\text{SO}(10)$ unification}
\label{sec:GCU}
As it is well known, in non-SUSY $\text{SO}(10)$ scenarios, the unification of the gauge couplings can be as good as, or
even better than, in the MSSM, despite that the number of extra fields up to the SM is small. This extra particle content can successfully fulfill all the constraints coming from the fermion masses, proton decay, and perturbativity.
In addition, if more restrictive conditions are imposed like a simplified DM model spectrum, the required extra field content needs to be more specific. 
In what follows, we will concentrate on these kinds of non-SUSY $\text{SO}(10)$ scenarios, focusing on two different channels to break $\text{SO}(10)$ to the SM, containing each one the remnant $U(1)_{B-L}$ symmetry necessary to stabilize DM.
The first scenario to analyze is based on the $\text{SO}(10)\rightarrow
\text{SU}(5)\times \text{U}(1)_{X}$ breaking channel.
Here, a PSS-like spectrum with 
singlet-doublet-triplet fermion DM is considered.
One well-known possibility in this chain is to have triplet fermion
dark matter at low energy with a fermion octect at one intermediate
scale. 
In order to have only singlet-doublet dark matter at low energies, we explore a 
second scenario  based in the left-right symmetry breaking chain. 
Very simple configurations of fields which not only explain rich
phenomenology but also DM through the singlet-doublet DM realization
are analyzed.

\subsection{Partial split supersymmetry-like model}

Here, we consider the symmetry breaking channel:
\begin{equation}
\text{SO}(10)\rightarrow \text{SU}(5)\times \text{U}(1)_{X}\rightarrow \text{SM}\,.
\end{equation}
In order to avoid intermediate breaking scales, we assume that $\text{SU}(5)\times \text{U}(1)_{X}$ breaks to the SM also at the unification scale  $m_{G}$, joint with the $\text{SO}(10)$ symmetry breaking.
At the first step, we are adding the two fermion doublets $\chi$,  $\chi^{c}$ at the electroweak scale $m_{\text{EW}}=100$ GeV. 
%Another way to have unification is by considering the adding to the spectrum~\eqref{eq:LNE2} the two fermion doublets $\chi$,  $\chi^{c}$ around the electroweak scale $m_{\text{EW}}=100$ GeV. 
At this scale an extra configuration of fields, denoted as $X$, is added such that $SM+\chi+\chi^{c}+X$ unifies equal or better than the MSSM at a scale of $m_{G}$. 
The $X$ configuration and the unification scale $m_{G}$ depend on the value of the new physics scale $m_{\text{NP}}$. 
As a first example, if we assume $m_{\text{NP}}=m_{\text{EW}}$, that is, the doublet fermions and the rest of fields are added at the electroweak scale, one of the simplest and interesting configurations found corresponds to $X=\Phi_{1,2,1/2}+2\Phi_{1,3,0}+2\Phi_{8,1,0}$ \cite{Parida:2011wh} which unifies at a scale of $m_{G}=2\times 10^{16}$ GeV,  when $M_D=100\ $GeV. A more general scan is to be presented below.
This configuration is also denoted in the literature as $\Phi_{1,2,1/2}+\Psi_{1,3,0}+\Psi_{8,1,0}$ since in our case two scalar fields, $2\Phi$, correspond to one fermionic field $\Psi$. 
Note that $X=\Phi_{1,2,1/2}+\Psi_{1,3,0}+\Psi_{8,1,0}+N$  (with two $45_F$, see Eq. (\ref{eq:LNE2})) is the same split-SUSY configuration but with the second scalar doublet ($\Phi_{1,2,1/2}$) living at low energies, which has been dubbed as partial split-SUSY~\cite{Diaz:2006ee}. 
Furthermore, $X+\chi+\chi^{c}$ has the same fermion fields of the
singlet-doublet-triplet fermion DM scenario discussed in the previous
section. 
Therefore, that $\text{SO}(10)$-based scenario is compatible with gauge
coupling unification.  
Of course, the DTFDM model is also compatible with GCU since the SM
singlets have no impact on it. 
With regard to the additional scalar doublet, if it proceeded from the
${\bf 16}_H$ representation, which is $P_M$-odd, and did not develop a
VEV, then the dark matter stability would be still guaranteed by the
matter parity as first noted in~\cite{Kadastik:2009dj,Kadastik:2009cu}\footnote{In other words, the second scalar doublet would
  be an inert doublet
  \cite{Deshpande:1977rw,Barbieri:2006dq,LopezHonorez:2006gr}.}. 
Moreover, with the $\Psi_{1,3,0}$ and $N$ from the two $\mathbf{45}_F$ and $\Phi_{1,2,1/2}$ from the $\mathbf{16}_{H}$, it is possible to build a hybrid  type-III~\cite{Ma:2008cu} and
type-I~\cite{Ma:2006km} radiative seesaw in $\text{SO}(10)$ as analyzed in~\cite{Parida:2011wh}.  

%
%However, when $m_{\Sigma}\sim m_G$, eq. \eqref{eq:NE} can be used to have one innocuous $\mathbf{45}_F$ with a free $m_{N}$  not affecting unification.  {\color{blue} I DON'T UNDERSTAND}
%

\begin{figure}
  \centering
\includegraphics[scale=0.53]{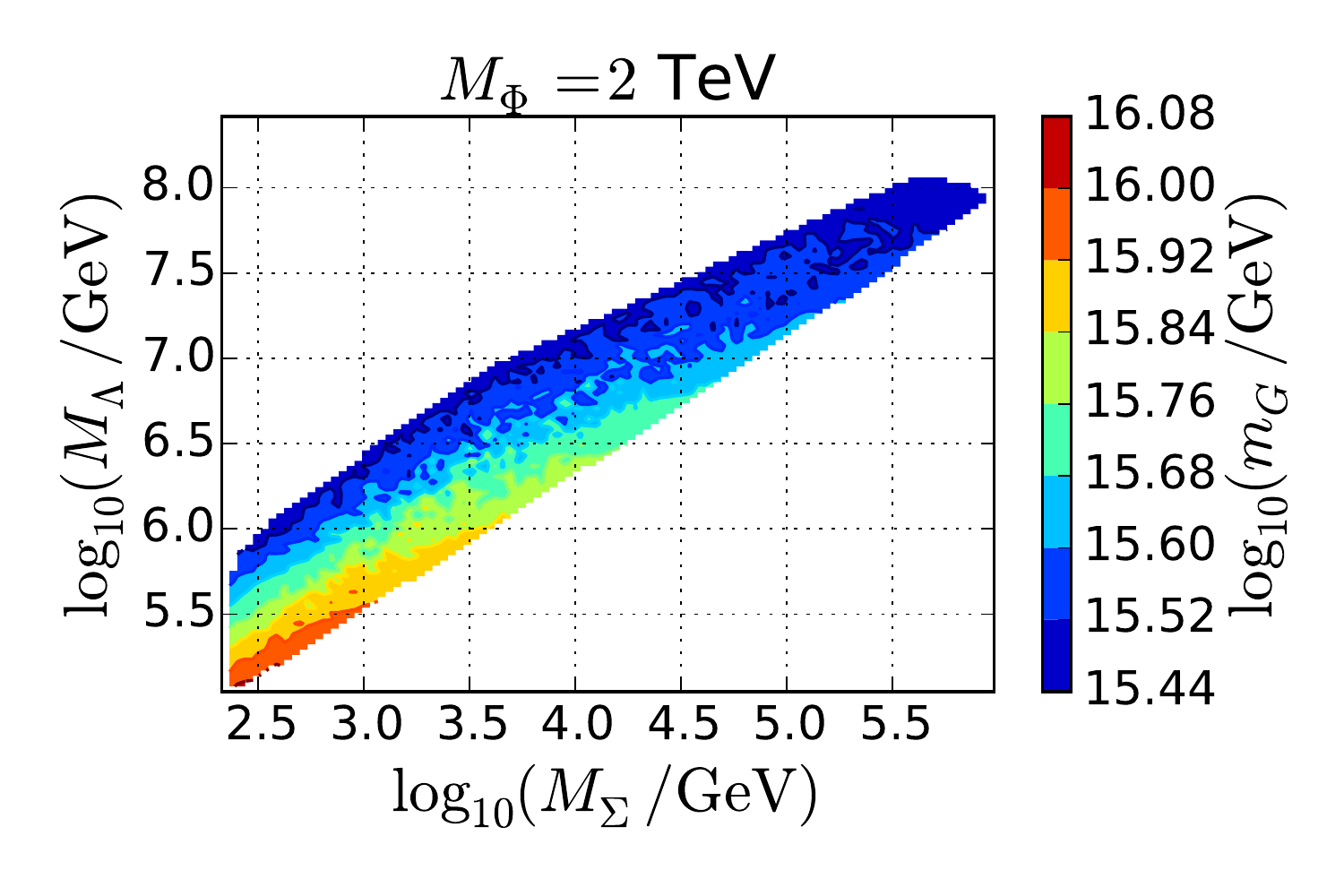}   \includegraphics[scale=0.53]{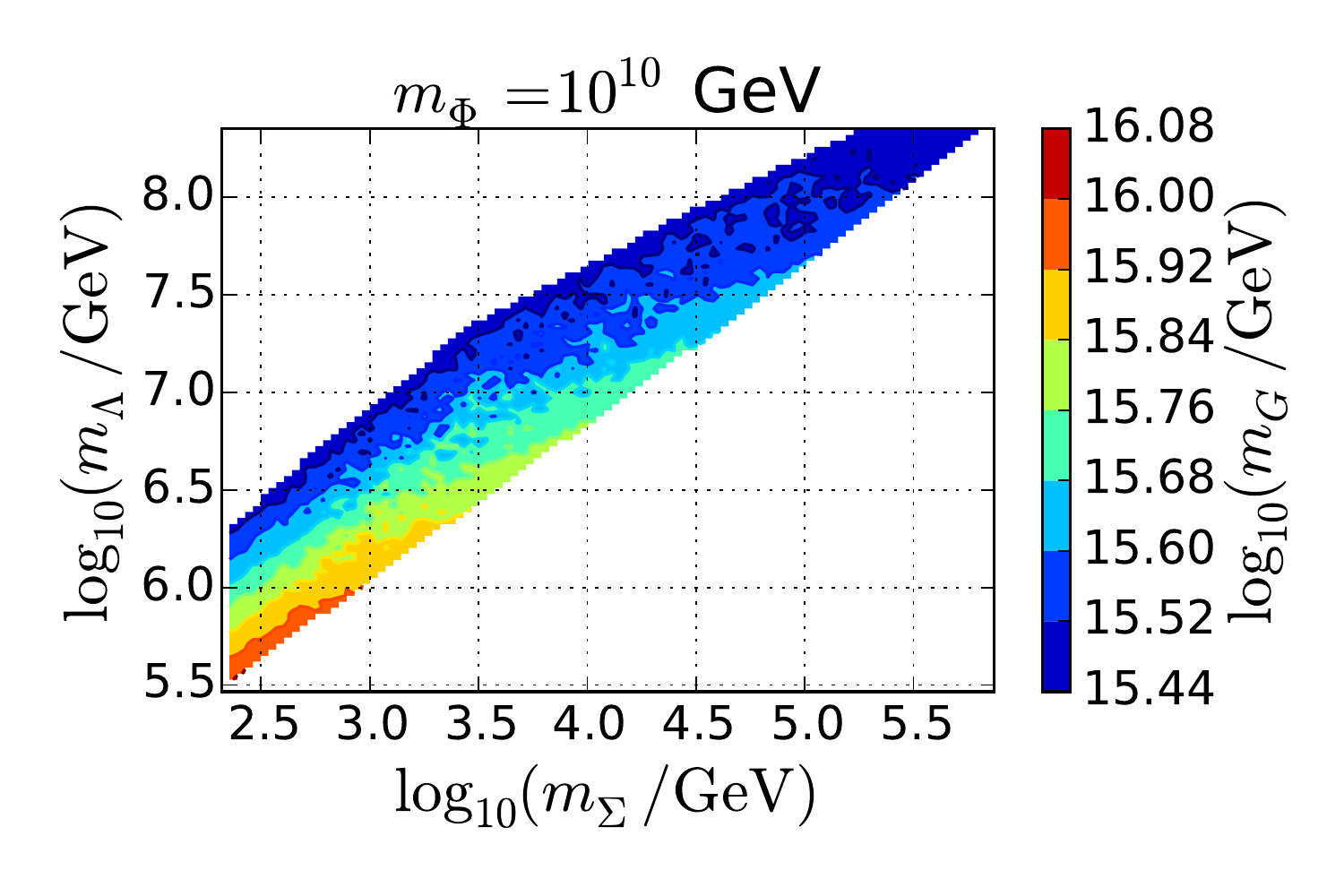}  %\\
%(a)\hfill (b)
  \caption{$SU(2)$-triplet fermion mass as a function of the $SU(3)$-octet fermion mass for $M_\Phi=2$\ TeV (left panel), and $M_\Phi=10^{10}$\ GeV (right panel). The red (blue) colors in the lower (upper) part of the region signal for high (low) unification scales compatible with proton decays.}
  \label{fig:one45}
\end{figure}

However, there is a more economical possibility by using a single
$\mathbf{45}_F$ evolving the spectrum~\eqref{eq:LNE}. In that case we need to
consider the effect of the $T$ color triplet in the running of the renormalization group equations (RGEs). To study the possibility of unification scale in that case, we scan the parameter space with
\begin{align}
0\le M_N/\text{GeV}\le& 3000\,, &  100\le &M_D/\text{GeV}\le 3000\,,
\end{align}
with either $M_\Sigma>\min(M_N,M_D)$  or  $M_\Sigma=2.7\ \text{TeV}$. The lightest neutral eigenvalue from the mass matrix in Eq.~\eqref{eq:Mchi} is checked to have the proper dark matter relic density and compatibility with all the phenomenological constraints as explained in~Ref.~\cite{Calibbi:2015nha}.  For each point in the scan we check if it is possible to choose
 $M_\Lambda$ and $M_\Phi$ to get proper unification within the range $3\times 10^{15}<m_{G}/\text{GeV}<10^{18}$. The results are shown in Fig.~\ref{fig:one45}. There we show $M_{\Lambda}$ as a function of $M_{\Sigma}$ for $M_{\Phi}=m_{\Phi(1,2,1/2)}=2\ \text{TeV}$ (left panel) and $M_{\Phi}=10^{10}\ \text{GeV}$ (right panel). In both figures, the several colors represent the possible values of $m_{G}$, ranging from the dark-blue color for $m_G\approx 3\times 10^{15}\ \text{GeV}$ to the red color for $m_G\approx 1.2\times 10^{16}\ \text{GeV}$. In this way, large unification scales are obtained for low $M_{\Sigma}$ and $M_{\Lambda}$, with a minimum value of $M_{\Lambda}$ around $100\ \text{TeV}$ for $M_{\Phi}=2\ \text{TeV}$ and $300\ \text{TeV}$ for $M_{\Phi}=10^{10}\ \text{GeV}$. We can see that the effect of the doublet scalar is to rescale the mass of the color octet with a factor of $3$ for their quoted values. Moreover, the results have only a mild dependence on the specific choice of $M_D$ and $M_N$ when the RGEs at one-loop are used to analyze unification.

\begin{figure}
  \centering
\includegraphics[scale=0.53]{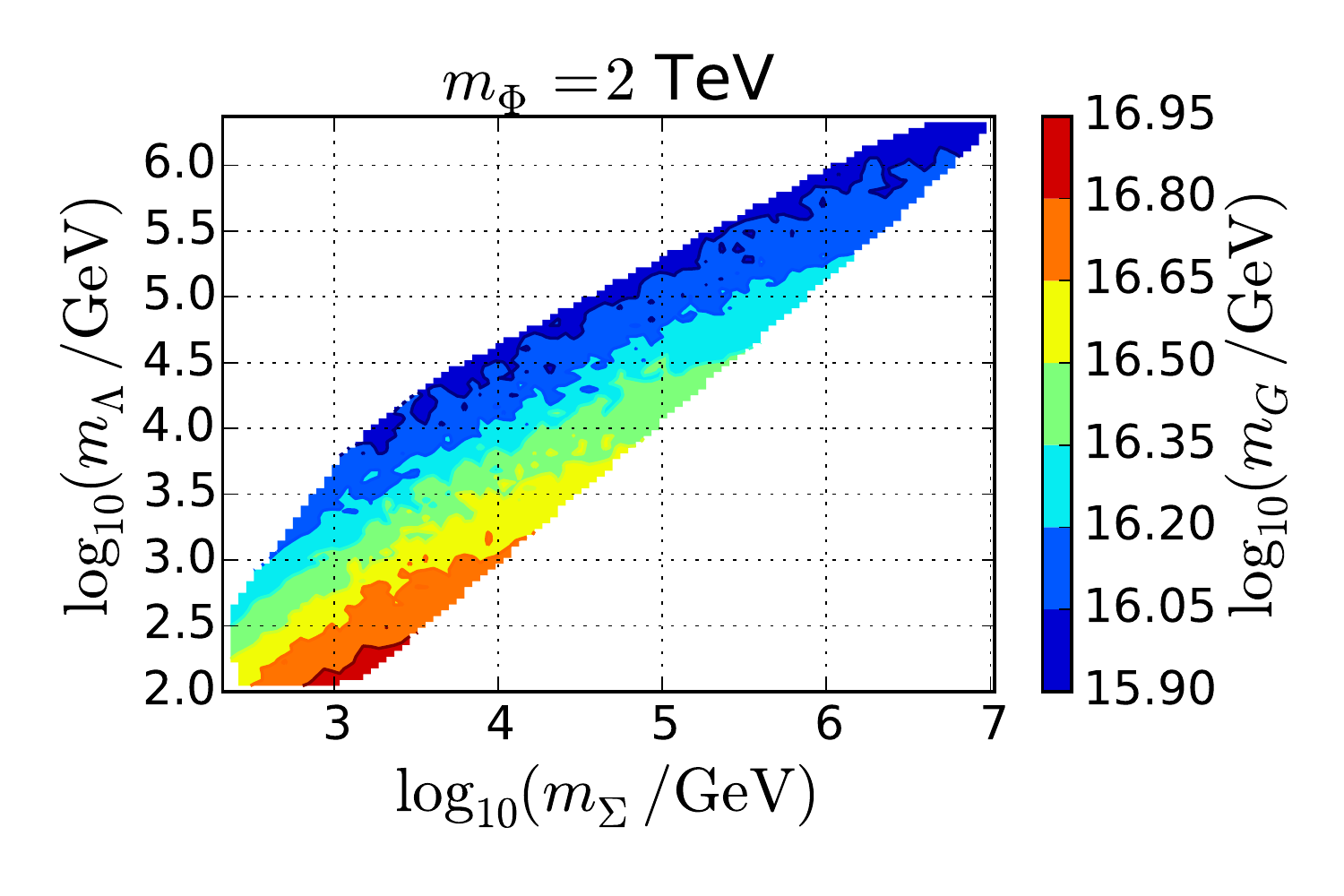}  
  \caption{The same as in left panel of Fig.~\ref{fig:one45}, but with two $\mathbf{45}_F$}
  \label{fig:two45}
\end{figure}

For completeness, we also show the lower $M_{\Lambda}$ scale allowed in the case of two $\mathbf{45}_F$ in Fig.~\ref{fig:two45} when $M_{\Phi}=2\ \text{TeV}$. We can see that the unification scale at $m_{G}=2\times 10^{16}\ \text{GeV}$ can be obtained from close to electroweak scale values for $M_{\Sigma}$ ($M_{\Lambda}$), until  $M_{\Sigma}\lesssim 1000\ \text{TeV}$ $(M_{\Lambda}\lesssim 100\ \text{TeV})$.

The required colored octet has been shown in~\cite{Frigerio:2009wf,Parida:2011wh}, to have an abundance and lifetime sufficiently small to satisfy all
experimental constraints~\cite{ArkaniHamed:2004fb}. In particular,  if the lifetime of the octet is long enough,
this would hadronize into R-hadrons as in split-SUSY, and the limits from ATLAS~\cite{Aad:2013gva} or CMS~\cite{Khachatryan:2015jha} for this kind of states would apply. 
In the later experimental study,  a colored octet with mass less than $880\,\text{GeV}$ is excluded if it decays into a gluon and the DM fermion candidate with a branching of 100\% and a lifetime between $1\,\mu\text{s}$  and $1000\,\textrm{s}$, providing that  the energy of the final gluon is larger than $120\,\text{GeV}$.
% * <restrepo@udea.edu.co> 2015-10-01T21:56:41.596Z:
%
% 
%

The fermion spectrum given~\eqref{eq:LE} was used in Ref.~\cite{Hambye:2009pw} where a  DM scheme arise from a simple unification configuration
containing only a fermion DM triplet at low energies at the price of
having a fermion octet at high energies~\cite{Hambye:2009pw} (the
Dirac fermion case is analyzed  in \cite{vanderBij:2012ck}). 
To have the proper DM relic abundance with a fermion-triplet of
$2.7$~TeV, the fermion octet needs to have a mass in a narrow range around
$2\times 10^{10}$~GeV. 

We now check if it is possible to have a pure SDFDM realization with
an intermediate left-right symmetry scale.

\subsection{Breaking through left-right chain}

As previously motivated, this scenario represents another possibility
to link the DM with GCU. 
In this case, we will concentrate only in the singlet-doublet fermion DM  scenario,
showing some simple LR configuration of fields which fulfill some
other phenomenological requirements.

For this model construction, we consider a chain in which $\text{SO}(10)$ is broken in exactly one intermediate LR step to the standard model group as:
\begin{equation}
\text{SO}(10)\rightarrow \text{SU}(3)_{c}\times \text{SU}(2)_{L}\times \text{SU}(2)_{R}\times \text{U}(1)_{B-L}\rightarrow \text{SM}\,.
\end{equation}
The left-right symmetry breaking scale (denoted in this case as
$m_{\text{LR}}$) can be as low as $O(1)$ TeV or
as high as, said $10^{9}$ GeV, maintaining nevertheless gauge coupling
unification  into the scheme of $\text{SO}(10)$. 
In Refs. \cite{DeRomeri:2011ie,Arbelaez:2013hr,Arbelaez:2013nga}
$\text{SO}(10)$ models with an intermediate left-right symmetry have been
studied. There, simple configurations which unify and also contain some
interesting phenomenological aspects have been explored.

For the construction of our configurations, we have taken the basic particle content described in Table~\ref{tab:ModelI} which consists of the SM fermions, plus the SM Higgs ($\Phi$), and the new particle content of the SDFDM.

%\begin{table}[t,floatfix]
\begin{table}[t]
\begin{tabular}{|c|c|c|c|c|}
\hline
{Field} & Multiplicity & $3_{c}2_{L}2_{R}1_{B-L}$& Spin & $\text{SO}(10)$ origin\\
\hline
$Q$ & 3 & $(3,2,1,+\tfrac{1}{3})$ & 1/2 &16\\
$Q^{c}$ & 3 & $(\bar 3,1,2,-\tfrac{1}{3})$ &1/2& 16\\
$L$  & 3 & $(1,2,1,-1)$& 1/2& 16\\
$L^{c}$& 3 & $(1,1,2,+1)$  & 1/2& 16\\
\hline
$\Phi$ & 1 & $(1,2,2,0)$ & 0& $10$\\
$\chi$, $\chi^{c}$ & 1 & $(1,2,2,0)$ %$\oplus (1,2,1,-1)$ 
 & 1/2& $10$\\
 $N$  & 1 & $(1,1,1,0)$& 1/2& 45\\
\hline 
\end{tabular}
\caption{The relevant part of the field content. Note that, the two fermion doublets $\chi$ and $\chi^{c}$ come from an only fermionic LR bidoublet. In the third column the relevant fields are
characterized by their $\text{SU}(3)_{c}\times \text{SU}(2)_{L}\times
\text{SU}(2)_{R}\times \text{U}(1)_{B-L}$ quantum numbers while their $\text{SO}(10)$
origin is specified in the fourth column. }
\label{tab:ModelI}
\end{table}

Therefore, at this point, the $\beta$-function contributions of the basic fields in Table~\ref{tab:ModelI} for the two regimes $[m_{\text{EW}},m_{\text{LR}}]$ and $[m_{\text{LR}},m_{G}]$
are given as:
\begin{align}
(b_{3}^{\text{SM}}, b_{2}^{\text{SM}}, b_{1}^{\text{SM}})&=(-7+\Delta b^{DM}_{3}, -19/6+\Delta b^{DM}_{2} ,41/10+\Delta b^{DM}_{1}), \nonumber \\
(b_{3}^{\text{LR}}, b_{2}^{\text{LR}}, b_{R}^{\text{LR}}, b_{B-L}^{\text{LR}})&=(-7,-7/3,-7/3,4)+(\Delta b_{3}^{\text{LR}}, \Delta b_{2}^{\text{LR}}, \Delta b_{R}^{\text{LR}}, \Delta b_{B-L}^{\text{LR}})\,,
\end{align}
where the $\Delta b^{DM}_{i}=(0,2/3,2/5)$ are the contributions of the two additional fermion DM doublets $\chi$ and $\chi^{c}$. We are using the canonical~($C$) normalization for the $(B-L)$ charge related to the physical~($P$) one, by $(B-L)^{C}=\sqrt{{3}/{8}}(B-L)^{P}$. Here, the $\Delta b_{i}^{\text{LR}}$ stands for the contribution of the additional fields which are added at the LR intermediate scale. It is clear that after adding only two fermionic doublets to the SM particle content, the gauge couplings
still do not unify. 
Only once the additional fields are added at the LR regime, unification is achieved at a scale of about $[10^{15}, 10^{17}]$ GeV (fulfilling this the actual  bounds that proton decay imposes in the GUT scale). 
As previously mentioned, $\chi$ and $\chi^{c}$ are added at the SM scale, so the interactions of DM with the LR gauge bosons and the other particles in this regime do not arise in this scenario.
To construct our models, besides to impose the presence of $\chi$, and $\chi^{c}$ and $N$ at the SM level, we require also a number of additional conditions for a model to be both realistic and phenomenological interesting: (i)~all models must have the agents to break the LR symmetry to the SM group (this achieved by the field $\Phi_{1,1,3,-2}$),
(ii) all models must contain (at least) one of the minimal ingredients to generate a realistic CKM in the quark sector, i.e, at least one copy of the $\Phi_{1,2,2,0}$ bidoublet and a copy of the $\Phi_{1,1,3,0}$ right triplet, iii)~models must have perturbative gauge couplings, and (iv) $m_{G}$ should be large enough to prevent too rapid proton decay, i.e, $m_{G}\ge 3\times 10^{15}$~GeV~\cite{Abe:2013lua}.

Note that $m_{\text{LR}}$ should be low enough so that the new fields can be accessible at the experiment, similar to the analysis already done in ~\cite{Arbelaez:2013nga}. However, for completeness,  we will show the simplest configurations of field even for large values of $m_{\text{LR}}$ as is depicted in Table~\ref{tab:I}.

\begin{figure}
\includegraphics[scale=0.7]{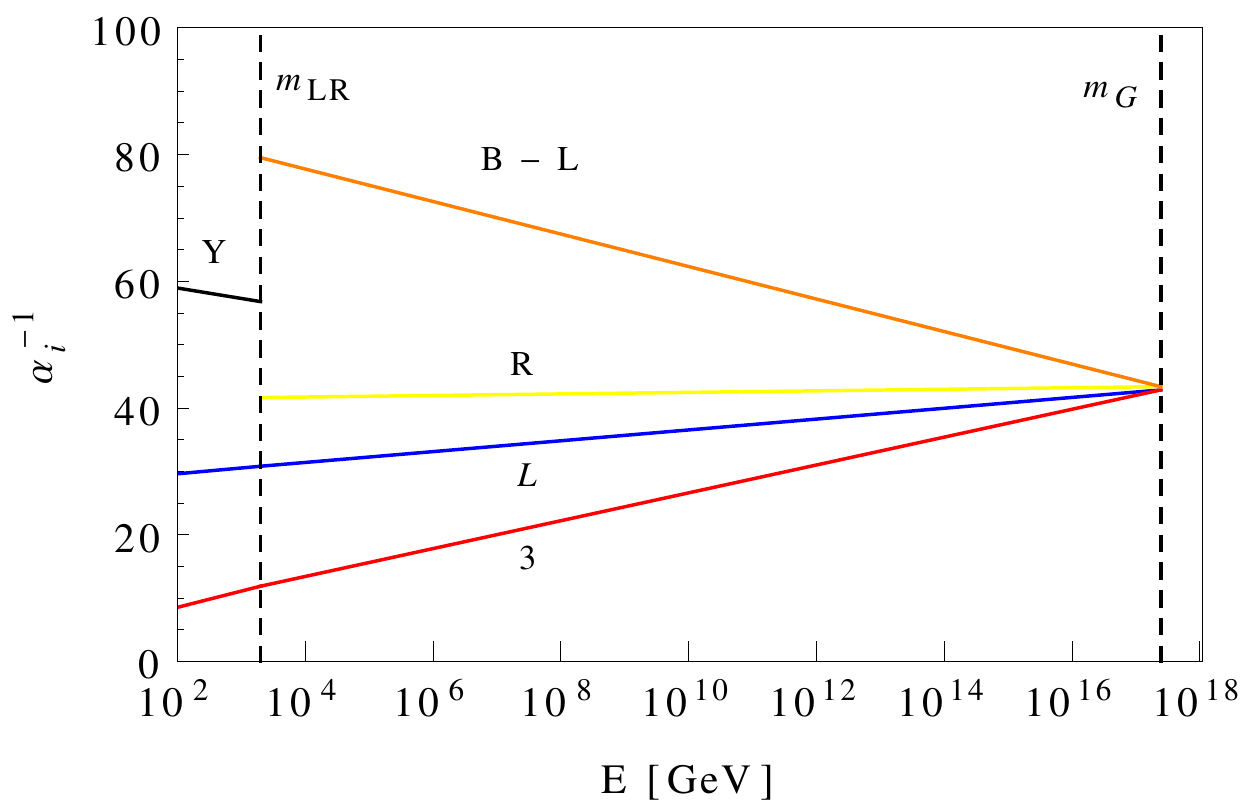}
\caption{Running of the gauge couplings for the first simple configuration shown in Table~\ref{tab:I}. In the regime $[m_{\text{EW}},m_{\text{LR}}]$ live the fields $SM+\Phi_{1,2,1/2}+\chi_{1,2,1/2}+\chi^{c}_{1,2,-1/2}+N_{1,1,0}$. In the second $[m_{\text{LR}},m_{G}]$ regime contribute the basic field depicted in Table~\ref{tab:ModelI} plus the extra fields: $\Phi_{1,1,3,0}+\Phi_{8,1,1,0}+2\Phi_{1,1,3,-2}$.}
\label{fig:1}
\end{figure}

The simplest of all the benchmark models passing the unification conditions above mentioned, with a left-right scale significantly low ($m_{\text{LR}}=2$~TeV) is  $\Phi_{1,1,3,0}+\Phi_{8,1,1,0}+2\Phi_{1,1,3,-2}$. Figure~\ref{fig:1} shows the running of the gauge couplings for this simple model. Note that all the fields in the LR regime are added at the LR scale of $2$~TeV. Considering this relatively low value of mass for the octet, there should be a chance for $\Phi_{8,1,0}$ to be within the reach of the current run of the LHC. The study of this scalar octet production has been already covered in the literature ~\cite{Bityukov:1997dh,Arnold:2011ra,Bai:2010dj}. At the LHC one of the contribution comes from the gluon-gluon annihilation into two scalar octets $gg\rightarrow \Phi_{8,1,0}$. For light scalar octets two gluon and quark-antiquark annihilations into two scalar octets give an additional contribution to the two-jet cross section ~\cite{Bityukov:1997dh}. There are also effects of the scalar octet on the process $pp\rightarrow 4$jets at the LHC, being this one of the best signatures to look for ~\cite{Arnold:2011ra}.
This is then an interesting solution that not only explains DM but also allows one to
explore rich phenomenology coming from the colored octet at the LHC.

It is worth to stressing that a sufficiently low LR scale is
still compatible with the interpretation for the ATLAS diboson
excess~\cite{Aad:2015owa} as a possible $W_R$
resonance~\cite{Brehmer:2015cia}. This will be reconsidered in future works where a detailed analysis of the interaction in the LR regime, in particular the $W_{L}-W_{R}$ possible mixings would be done.

\begin{table}%[h]
\begin{center}
\scalebox{0.85}{
\begin{tabular}{|c|c|c|}
\hline
 $m_{\text{LR}}\,(\mbox{GeV})$ &  LR configuration  & $m_{G}\,(\mbox{GeV})$ \\ \hline

 \multirow{3} {1.5cm} {$2\times 10^{3}$ } & $\Phi_{1,1,3,0}+\Phi_{8,1,1,0}+2\Phi_{1,1,3,-2}$& $2.47\times 10^{17}$  \\ \cline{2-3}
 
 & $3\Phi_{1,1,3,0}+\Phi_{1,2,2,0}+\Phi_{8,1,1,0}+2\Phi_{1,1,3,-2}$ & $1.65 \times 10^{16}$   \\ \cline{2-3} 
 
 & $\Phi_{1,1,3,0}+3\Phi_{3,1,1,4/3}+2\Phi_{1,1,3,-2}$ & $5.02 \times 10^{15}$   \\ \hline
                                                      
 \multirow{3} {1.5cm} {$10^{5}$} & $\Phi_{1,1,3,0}+\Phi_{8,1,1,0}+2\Phi_{1,1,3,-2}$& $1.01\times 10^{17}$   \\ \cline{2-3}

  & $2\Phi_{1,1,3,0}+\Phi_{1,2,2,0}+\Phi_{8,1,1,0}+2\Phi_{1,1,3,-2}$ & $1.01 \times 10^{16}$   \\ \cline{2-3} 
  
 & $\Phi_{1,1,3,0}+3\Phi_{3,1,1,4/3}+2\Phi_{1,1,3,-2}$ & $3.67 \times 10^{15}$   \\ \hline

 \multirow{3} {1.5cm} {$10^{7}$} & $\Phi_{1,1,3,0}+\Phi_{8,1,1,0}+2\Phi_{1,1,3,-2}$& $3.55\times 10^{16}$   \\ \cline{2-3}

  & $\Phi_{1,1,3,0}+\Phi_{1,2,2,0}+\Phi_{8,1,1,0}+2\Phi_{1,1,3,-2}$ & $5.69 \times 10^{15}$   \\ \cline{2-3} 
  
 & $\Phi_{1,1,3,0}+2\Phi_{3,1,1,-2/3}+3\Phi_{3,1,1,4/3}+\Phi_{1,1,3,-2}$ & $5.69 \times 10^{15}$   \\ \hline
                                           
\end{tabular}} %}
\end{center}
\caption{Simple LR configurations passing the constraints explained in the text. One of the scalar bidoublets $\Phi$ is already considered in the basic field content $(SM+\Phi+\chi+\chi^{c})$ as shown in Table~\ref{tab:ModelI}. The first configuration corresponds to the minimal solution. Both scales $m_{\text{LR}}$ and $m_{G}$ are given in GeV.}
\label{tab:I}
\end{table}
% * <restrepo@udea.edu.co> 2015-09-09T18:40:38.741Z:
%
%  3-2-2-1         SO(10)    4-2-2     4    3(1)
% \Phi(1,1,3,-2) -> 126_H-> (10,1,3): 10 -> 1(2)
% \Phi(1,2,2,0) -> 10_H  -> (1,2,2):   1 -> 1(0)
% \Phi(1,1,3,0) -> 45_H  -> (1,1,3):   1 -> 1(0)
% \Phi(8,1,1,0) -> 45_H  -> (15,1,1): 15 -> 8(0)
% \Phi(3,1,1,-2/3)->10_H -> (6,1,1):   6 -> 3(2/3)
% \Phi(3,1,1,-4/3)->45_H -> (15,1,1): 15 -> 3(4/3)
%

It is interesting to note that in the standard left-right symmetric models (without considering any extra particle contribution in the regime $[m_{\text{EW}},m_{\text{LR}}]$), the extra degree of freedom of having one intermediate $m_{\text{LR}}$ scale, allows one to achieve gauge coupling unification even if the extra particle content does not contain colored fields. However in our models, before reaching the LR intermediate stage, there is a previous $\Delta b_{2L}$ contribution which comes from the fermion doublets $\chi$ and $\chi^{c}$ already added at the SM scale, then some colored fields must be added at the LR scale in order to compensate this amount. On the other hand, if the scale at which the fermionic DM candidates are added is greater than $m_{\text{EW}}$ and very close to $m_{\text{LR}}$ it could be possible to obtain GCU without colored fields in the LR regime, but, all these solutions are excluded since  the unification scale is very low, i.e, $m_{G}< 3\times 10^{15}$ GeV.

\section{Conclusions}
\label{sec:conclusions}
In this paper  we have taken advantage of the fact that a $Z_2$ symmetry appears as a remnant symmetry at the end of the symmetry breaking chain of the $\text{SO}(10)$ GUT group to the standard model, for constructing simplified fermion dark matter models where the dark matter stability is naturally guaranteed.

Concretely, we have formulated a viable $\text{SO}(10)$ model capable of
realizing at low energies the singlet-doublet-triplet fermion dark
matter. The model engages as nonstandard fermions a SM singlet and a hyperchargeless weak triplet, both belonging to the ${\bf 45}_F$, 
and a couple of weak doublets with $Y=\pm1/2$ belonging to the ${\bf 10}_F$ representation. 
The  mixing between these fermions is carried out by the SM Higgs, which is assigned to the ${\bf 10}_H$ representation. 
At low energies the resulting particle spectrum resembles the neutralino and chargino sets of the MSSM but with the difference that the mixing terms are not controlled by the gauge couplings. 
Thanks to the versatility of the model, it is also possible to realize the simpler fermion DM scenarios of singlet-doublet, doublet-triplet or only triplet.
 
Regarding gauge coupling unification, we have shown that the model has a successful $\text{SO}(10)$ unification through the $\text{SU}(5)\times U(1)_{X}$ chain by requiring the additional presence of a scalar weak doublet and a fermion color octect. The DTFDM model shares this same feature while the SDFDM model requires the presence of other DM fields.  However, this model under the left-right symmetry intermediate chain successfully achieves $\text{SO}(10)$ unification by demanding only the exotic scalar representations associated to the breaking of the left-right symmetry. 

In summary, an interesting configuration of fields which pass some physical
conditions such as a GCU, proton stability, compatibility with the
quark and lepton masses and mixings, fermionic DM realization and some
other nontrivial LHC phenomenology were found for the two
$\text{SO}(10)$ breaking channels explored. For both cases, the new
extra particle content close to the TeV scale, would make the new
physics testable at the LHC test. 

\emph{Note added}: Recently, we became aware of the work of N. Nagata, K.A.Olive and J. Zheng~\cite{Nagata:2015dma} where singlet-doublet fermion dark matter model is studied in the limit where the singlet is at some high intermediate scale into the framework of $\text{SO}(10)$ unification. 

\section*{Acknowledgments}

C.A. acknowledges support by Fondecyt (Chile), Grant No. 3150472. D.R. and O.Z. have been partially supported by UdeA through the grants 
Sostenibilidad-GFIF, CODI-2014-361 and CODI-IN650CE, and COLCIENCIAS
through the Grants No. 111-556-934918 and No. 111-565-842691. R.L. is supported by COLCIENCIAS and acknowledges the hospitality of Universidade Federal do ABC while this work was being completed. We are grateful to Martin Hirsch for reading through the preliminary versions of the manuscript.

%%%%%%%%%%%%%%%%%%%%%%%%%%%%%%%%%%%%%%%%%%%
\bibliographystyle{h-physrev4}
\bibliography{darkmatter}

\begin{thebibliography}{10}

\bibitem{ArkaniHamed:2004fb}
N.~Arkani-Hamed and S.~Dimopoulos,
\newblock JHEP {\bf 06}, 073 (2005), arXiv:hep-th/0405159.
%%CITATION = HEP-TH/0405159;%%

\bibitem{Giudice:2004tc}
G.~F. Giudice and A.~Romanino,
\newblock Nucl. Phys. {\bf B699}, 65 (2004), arXiv:hep-ph/0406088,
\newblock [Erratum: Nucl. Phys.B706,65(2005)].
%%CITATION = HEP-PH/0406088;%%

\bibitem{ArkaniHamed:2004yi}
N.~Arkani-Hamed, S.~Dimopoulos, G.~F. Giudice, and A.~Romanino,
\newblock Nucl. Phys. {\bf B709}, 3 (2005), arXiv:hep-ph/0409232.
%%CITATION = HEP-PH/0409232;%%

\bibitem{Ibanez:1991hv}
L.~E. Ibanez and G.~G. Ross,
\newblock Phys. Lett. {\bf B260}, 291 (1991).
%%CITATION = PHLTA,B260,291;%%

\bibitem{Ibanez:1991pr}
L.~E. Ibanez and G.~G. Ross,
\newblock Nucl. Phys. {\bf B368}, 3 (1992).
%%CITATION = NUPHA,B368,3;%%

\bibitem{Babu:2003qh}
K.~S. Babu, I.~Gogoladze, and K.~Wang,
\newblock Phys. Lett. {\bf B570}, 32 (2003), arXiv:hep-ph/0306003.
%%CITATION = HEP-PH/0306003;%%

\bibitem{Dreiner:2005rd}
H.~K. Dreiner, C.~Luhn, and M.~Thormeier,
\newblock Phys. Rev. {\bf D73}, 075007 (2006), arXiv:hep-ph/0512163.
%%CITATION = HEP-PH/0512163;%%

\bibitem{Paraskevas:2012kn}
M.~Paraskevas and K.~Tamvakis,
\newblock Phys. Rev. {\bf D86}, 015009 (2012), arXiv:1205.1391.
%%CITATION = ARXIV:1205.1391;%%

\bibitem{Krauss:1988zc}
L.~M. Krauss and F.~Wilczek,
\newblock Phys. Rev. Lett. {\bf 62}, 1221 (1989).
%%CITATION = PRLTA,62,1221;%%

\bibitem{Banks:1991xj}
T.~Banks and M.~Dine,
\newblock Phys. Rev. {\bf D45}, 1424 (1992), arXiv:hep-th/9109045.
%%CITATION = HEP-TH/9109045;%%

\bibitem{Mambrini:2015sia}
Y.~Mambrini, S.~Profumo, and F.~S. Queiroz,
\newblock (2015), arXiv:1508.06635.
%%CITATION = ARXIV:1508.06635;%%

\bibitem{Sierra:2009zq}
D.~Aristizabal~Sierra, D.~Restrepo, and O.~Zapata,
\newblock Phys. Rev. {\bf D80}, 055010 (2009), arXiv:0907.0682.
%%CITATION = ARXIV:0907.0682;%%

\bibitem{Dreiner:2012ae}
H.~K. Dreiner, M.~Hanussek, and C.~Luhn,
\newblock Phys. Rev. {\bf D86}, 055012 (2012), arXiv:1206.6305.
%%CITATION = ARXIV:1206.6305;%%

\bibitem{Florez:2013mxa}
A.~Florez, D.~Restrepo, M.~Velasquez, and O.~Zapata,
\newblock Phys. Rev. {\bf D87}, 095010 (2013), arXiv:1303.0278.
%%CITATION = ARXIV:1303.0278;%%

\bibitem{Wang:2015mea}
F.~Wang, W.~Wang, and J.~M. Yang,
\newblock JHEP {\bf 03}, 050 (2015), arXiv:1501.02906.
%%CITATION = ARXIV:1501.02906;%%

\bibitem{Dimopoulos:1981zb}
S.~Dimopoulos and H.~Georgi,
\newblock Nucl. Phys. {\bf B193}, 150 (1981).
%%CITATION = NUPHA,B193,150;%%

\bibitem{Dimopoulos:1981dw}
S.~Dimopoulos, S.~Raby, and F.~Wilczek,
\newblock Phys. Lett. {\bf B112}, 133 (1982).
%%CITATION = PHLTA,B112,133;%%

\bibitem{Fritzsch:1974nn}
H.~Fritzsch and P.~Minkowski,
\newblock Annals Phys. {\bf 93}, 193 (1975).
%%CITATION = APNYA,93,193;%%

\bibitem{Cirelli:2005uq}
M.~Cirelli, N.~Fornengo, and A.~Strumia,
\newblock Nucl. Phys. {\bf B753}, 178 (2006), arXiv:hep-ph/0512090.
%%CITATION = HEP-PH/0512090;%%

\bibitem{Kadastik:2009dj}
M.~Kadastik, K.~Kannike, and M.~Raidal,
\newblock Phys. Rev. {\bf D81}, 015002 (2010), arXiv:0903.2475.
%%CITATION = ARXIV:0903.2475;%%

\bibitem{Kadastik:2009cu}
M.~Kadastik, K.~Kannike, and M.~Raidal,
\newblock Phys. Rev. {\bf D80}, 085020 (2009), arXiv:0907.1894,
\newblock [Erratum: Phys. Rev.D81,029903(2010)].
%%CITATION = ARXIV:0907.1894;%%

\bibitem{Frigerio:2009wf}
M.~Frigerio and T.~Hambye,
\newblock Phys. Rev. {\bf D81}, 075002 (2010), arXiv:0912.1545.
%%CITATION = ARXIV:0912.1545;%%

\bibitem{Parida:2010jj}
M.~K. Parida, P.~K. Sahu, and K.~Bora,
\newblock Phys. Rev. {\bf D83}, 093004 (2011), arXiv:1011.4577.
%%CITATION = ARXIV:1011.4577;%%

\bibitem{Mambrini:2015vna}
Y.~Mambrini, N.~Nagata, K.~A. Olive, J.~Quevillon, and J.~Zheng,
\newblock Phys. Rev. {\bf D91}, 095010 (2015), arXiv:1502.06929.
%%CITATION = ARXIV:1502.06929;%%

\bibitem{Parida:2011wh}
M.~K. Parida,
\newblock Phys. Lett. {\bf B704}, 206 (2011), arXiv:1106.4137.
%%CITATION = ARXIV:1106.4137;%%

\bibitem{Diaz:2006ee}
M.~A. Diaz, P.~Fileviez~Perez, and C.~Mora,
\newblock Phys. Rev. {\bf D79}, 013005 (2009), arXiv:hep-ph/0605285.
%%CITATION = HEP-PH/0605285;%%

\bibitem{Ma:2005he}
E.~Ma,
\newblock Phys. Lett. {\bf B625}, 76 (2005), arXiv:hep-ph/0508030.
%%CITATION = HEP-PH/0508030;%%

\bibitem{Patt:2006fw}
B.~Patt and F.~Wilczek,
\newblock (2006), arXiv:hep-ph/0605188.
%%CITATION = HEP-PH/0605188;%%

\bibitem{ArkaniHamed:2005yv}
N.~Arkani-Hamed, S.~Dimopoulos, and S.~Kachru,
\newblock (2005), arXiv:hep-th/0501082.
%%CITATION = HEP-TH/0501082;%%

\bibitem{Mahbubani:2005pt}
R.~Mahbubani and L.~Senatore,
\newblock Phys.Rev. {\bf D73}, 043510 (2006), arXiv:hep-ph/0510064.
%%CITATION = HEP-PH/0510064;%%

\bibitem{D'Eramo:2007ga}
F.~D'Eramo,
\newblock Phys.Rev. {\bf D76}, 083522 (2007), arXiv:0705.4493.
%%CITATION = ARXIV:0705.4493;%%

\bibitem{Enberg:2007rp}
R.~Enberg, P.~Fox, L.~Hall, A.~Papaioannou, and M.~Papucci,
\newblock JHEP {\bf 0711}, 014 (2007), arXiv:0706.0918.
%%CITATION = ARXIV:0706.0918;%%

\bibitem{Dedes:2014hga}
A.~Dedes and D.~Karamitros,
\newblock Phys.Rev. {\bf D89}, 115002 (2014), arXiv:1403.7744.
%%CITATION = ARXIV:1403.7744;%%

\bibitem{Djouadi:2011aa}
A.~Djouadi, O.~Lebedev, Y.~Mambrini, and J.~Quevillon,
\newblock Phys. Lett. {\bf B709}, 65 (2012), arXiv:1112.3299.
%%CITATION = ARXIV:1112.3299;%%

\bibitem{Kim:2006af}
Y.~G. Kim and K.~Y. Lee,
\newblock Phys. Rev. {\bf D75}, 115012 (2007), arXiv:hep-ph/0611069.
%%CITATION = HEP-PH/0611069;%%

\bibitem{Kim:2008pp}
Y.~G. Kim, K.~Y. Lee, and S.~Shin,
\newblock JHEP {\bf 05}, 100 (2008), arXiv:0803.2932.
%%CITATION = ARXIV:0803.2932;%%

\bibitem{Baek:2011aa}
S.~Baek, P.~Ko, and W.-I. Park,
\newblock JHEP {\bf 02}, 047 (2012), arXiv:1112.1847.
%%CITATION = ARXIV:1112.1847;%%

\bibitem{LopezHonorez:2012kv}
L.~Lopez-Honorez, T.~Schwetz, and J.~Zupan,
\newblock Phys. Lett. {\bf B716}, 179 (2012), arXiv:1203.2064.
%%CITATION = ARXIV:1203.2064;%%

\bibitem{Fedderke:2014wda}
M.~A. Fedderke, J.-Y. Chen, E.~W. Kolb, and L.-T. Wang,
\newblock JHEP {\bf 08}, 122 (2014), arXiv:1404.2283.
%%CITATION = ARXIV:1404.2283;%%

\bibitem{Freitas:2015hsa}
A.~Freitas, S.~Westhoff, and J.~Zupan,
\newblock (2015), arXiv:1506.04149.
%%CITATION = ARXIV:1506.04149;%%

\bibitem{Siringo:2012bc}
F.~Siringo,
\newblock Phys. Part. Nucl. Lett. {\bf 10}, 94 (2013), arXiv:1208.3599.
%%CITATION = ARXIV:1208.3599;%%

\bibitem{Lindner:1996tf}
M.~Lindner and M.~Weiser,
\newblock Phys. Lett. {\bf B383}, 405 (1996), arXiv:hep-ph/9605353.
%%CITATION = HEP-PH/9605353;%%

\bibitem{Dev:2009aw}
P.~S.~B. Dev and R.~N. Mohapatra,
\newblock Phys. Rev. {\bf D81}, 013001 (2010), arXiv:0910.3924.
%%CITATION = ARXIV:0910.3924;%%

\bibitem{Bertolini:2009es}
S.~Bertolini, L.~Di~Luzio, and M.~Malinsky,
\newblock Phys. Rev. {\bf D81}, 035015 (2010), arXiv:0912.1796.
%%CITATION = ARXIV:0912.1796;%%

\bibitem{DeRomeri:2011ie}
V.~De~Romeri, M.~Hirsch, and M.~Malinsky,
\newblock Phys. Rev. {\bf D84}, 053012 (2011), arXiv:1107.3412.
%%CITATION = ARXIV:1107.3412;%%

\bibitem{Arbelaez:2013hr}
C.~Arbelaez, R.~M. Fonseca, M.~Hirsch, and J.~C. Romao,
\newblock Phys. Rev. {\bf D87}, 075010 (2013), arXiv:1301.6085.
%%CITATION = ARXIV:1301.6085;%%

\bibitem{Malinsky:2007qy}
M.~Malinsky,
\newblock Phys. Rev. {\bf D77}, 055016 (2008), arXiv:0710.0581.
%%CITATION = ARXIV:0710.0581;%%

\bibitem{Bajc:2005zf}
B.~Bajc, A.~Melfo, G.~Senjanovic, and F.~Vissani,
\newblock Phys. Rev. {\bf D73}, 055001 (2006), arXiv:hep-ph/0510139.
%%CITATION = HEP-PH/0510139;%%

\bibitem{Babu:2015bna}
K.~S. Babu and S.~Khan,
\newblock Phys. Rev. {\bf D92}, 075018 (2015), arXiv:1507.06712.
%%CITATION = ARXIV:1507.06712;%%

\bibitem{Martin:1997ns}
S.~P. Martin,
\newblock (1997), arXiv:hep-ph/9709356,
\newblock [Adv. Ser. Direct. High Energy Phys.18,1(1998)].
%%CITATION = HEP-PH/9709356;%%

\bibitem{Carena:2004ha}
M.~Carena, A.~Megevand, M.~Quiros, and C.~E.~M. Wagner,
\newblock Nucl. Phys. {\bf B716}, 319 (2005), arXiv:hep-ph/0410352.
%%CITATION = HEP-PH/0410352;%%

\bibitem{Benakli:2013msa}
K.~Benakli, L.~Darmé, M.~D. Goodsell, and P.~Slavich,
\newblock JHEP {\bf 05}, 113 (2014), arXiv:1312.5220.
%%CITATION = ARXIV:1312.5220;%%

\bibitem{Abe:2014gua}
T.~Abe, R.~Kitano, and R.~Sato,
\newblock (2014), arXiv:1411.1335.
%%CITATION = ARXIV:1411.1335;%%

\bibitem{Cohen:2011ec}
T.~Cohen, J.~Kearney, A.~Pierce, and D.~Tucker-Smith,
\newblock Phys.Rev. {\bf D85}, 075003 (2012), arXiv:1109.2604.
%%CITATION = ARXIV:1109.2604;%%

\bibitem{Cheung:2013dua}
C.~Cheung and D.~Sanford,
\newblock JCAP {\bf 1402}, 011 (2014), arXiv:1311.5896.
%%CITATION = ARXIV:1311.5896;%%

\bibitem{Restrepo:2015ura}
D.~Restrepo, A.~Rivera, M.~Sánchez-Peláez, O.~Zapata, and W.~Tangarife,
\newblock Phys. Rev. {\bf D92}, 013005 (2015), arXiv:1504.07892.
%%CITATION = ARXIV:1504.07892;%%

\bibitem{Calibbi:2015nha}
L.~Calibbi, A.~Mariotti, and P.~Tziveloglou,
\newblock (2015), arXiv:1505.03867.
%%CITATION = ARXIV:1505.03867;%%

\bibitem{Abdallah:2015ter}
J.~Abdallah {\em et~al.},
\newblock Phys. Dark Univ. {\bf 9-10}, 8 (2015), arXiv:1506.03116.
%%CITATION = ARXIV:1506.03116;%%

\bibitem{Chattopadhyay:2005mv}
U.~Chattopadhyay, D.~Choudhury, M.~Drees, P.~Konar, and D.~Roy,
\newblock Phys.Lett. {\bf B632}, 114 (2006), arXiv:hep-ph/0508098.
%%CITATION = HEP-PH/0508098;%%

\bibitem{Deshpande:1977rw}
N.~G. Deshpande and E.~Ma,
\newblock Phys.Rev. {\bf D18}, 2574 (1978).
%%CITATION = PHRVA,D18,2574;%%

\bibitem{Barbieri:2006dq}
R.~Barbieri, L.~J. Hall, and V.~S. Rychkov,
\newblock Phys.Rev. {\bf D74}, 015007 (2006), arXiv:hep-ph/0603188.
%%CITATION = HEP-PH/0603188;%%

\bibitem{LopezHonorez:2006gr}
L.~Lopez~Honorez, E.~Nezri, J.~F. Oliver, and M.~H.~G. Tytgat,
\newblock JCAP {\bf 0702}, 028 (2007), arXiv:hep-ph/0612275.
%%CITATION = HEP-PH/0612275;%%

\bibitem{Ma:2008cu}
E.~Ma and D.~Suematsu,
\newblock Mod. Phys. Lett. {\bf A24}, 583 (2009), arXiv:0809.0942.
%%CITATION = ARXIV:0809.0942;%%

\bibitem{Ma:2006km}
E.~Ma,
\newblock Phys.Rev. {\bf D73}, 077301 (2006), arXiv:hep-ph/0601225.
%%CITATION = HEP-PH/0601225;%%

\bibitem{Aad:2013gva}
ATLAS, G.~Aad {\em et~al.},
\newblock Phys. Rev. {\bf D88}, 112003 (2013), arXiv:1310.6584.
%%CITATION = ARXIV:1310.6584;%%

\bibitem{Khachatryan:2015jha}
CMS, V.~Khachatryan {\em et~al.},
\newblock Eur. Phys. J. {\bf C75}, 151 (2015), arXiv:1501.05603.
%%CITATION = ARXIV:1501.05603;%%

\bibitem{Hambye:2009pw}
T.~Hambye, F.~S. Ling, L.~Lopez~Honorez, and J.~Rocher,
\newblock JHEP {\bf 07}, 090 (2009), arXiv:0903.4010,
\newblock [Erratum: JHEP05,066(2010)].
%%CITATION = ARXIV:0903.4010;%%

\bibitem{vanderBij:2012ck}
J.~J. van~der Bij,
\newblock Europhys. Lett. {\bf 100}, 29003 (2012), arXiv:1207.5317.
%%CITATION = ARXIV:1207.5317;%%

\bibitem{Arbelaez:2013nga}
C.~Arbeláez, M.~Hirsch, M.~Malinský, and J.~C. Romão,
\newblock Phys. Rev. {\bf D89}, 035002 (2014), arXiv:1311.3228.
%%CITATION = ARXIV:1311.3228;%%

\bibitem{Abe:2013lua}
Super-Kamiokande, K.~Abe {\em et~al.},
\newblock Phys. Rev. Lett. {\bf 113}, 121802 (2014), arXiv:1305.4391.
%%CITATION = ARXIV:1305.4391;%%

\bibitem{Bityukov:1997dh}
S.~I. Bityukov and N.~V. Krasnikov,
\newblock Mod. Phys. Lett. {\bf A12}, 2011 (1997), arXiv:hep-ph/9705338.
%%CITATION = HEP-PH/9705338;%%

\bibitem{Arnold:2011ra}
J.~M. Arnold and B.~Fornal,
\newblock Phys. Rev. {\bf D85}, 055020 (2012), arXiv:1112.0003.
%%CITATION = ARXIV:1112.0003;%%

\bibitem{Bai:2010dj}
Y.~Bai and B.~A. Dobrescu,
\newblock JHEP {\bf 07}, 100 (2011), arXiv:1012.5814.
%%CITATION = ARXIV:1012.5814;%%

\bibitem{Aad:2015owa}
ATLAS, G.~Aad {\em et~al.},
\newblock (2015), arXiv:1506.00962.
%%CITATION = ARXIV:1506.00962;%%

\bibitem{Brehmer:2015cia}
J.~Brehmer, J.~Hewett, J.~Kopp, T.~Rizzo, and J.~Tattersall,
\newblock (2015), arXiv:1507.00013.
%%CITATION = ARXIV:1507.00013;%%

\bibitem{Nagata:2015dma}
N.~Nagata, K.~A. Olive, and J.~Zheng,
\newblock (2015), arXiv:1509.00809.
%%CITATION = ARXIV:1509.00809;%%

\end{thebibliography}
%%%%%%%%%%%%%%%%%%%%%%%%%%%%%%%%%%%%%%%%%%%%%%%%%

\end{document}